\documentclass{caosp309}

\usepackage{graphicx}
\usepackage{color}

\usepackage{natbib}
\articleNo{301}
\pubyear{2025}
\volume{55}
\volnumber{3}
\firstpage{452}
\received{October 29, 2024}
\accepted{February 5, 2025}

\def\BibTeX{{\rm B\kern-.05em{\sc i\kern-.025em b}\kern-.08em
             T\kern-.1667em\lower.7ex\hbox{E}\kern-.125emX}}

\begin{document}

%
\hauthor{C.I.\,Eze, G.\,Handler, F.Kahraman \,Ali\c{c}avu\c{s}, T.\,Pawar and A.\,Miszuda}

\title{Characterizing the variability of a sample of massive stars in eclipsing binaries}



\author{
        C.I.\,Eze\inst{1,2,*}\orcid{0000-0003-3119-0399}
      \and
        G.\,Handler\inst{1}\orcid{0000-0001-7756-1568} 
      \and 
        F.Kahraman\,Ali\c{c}avu\c{s}\inst{3}\orcid{0000-0002-9036-7476}   
      \and 
        T.\,Pawar\inst{4}\orcid{0000-0002-0004-0569}
      \and 
        A.\,Miszuda\inst{1}\orcid{0000-0002-9382-2542}
       }


\institute{
           Nicolaus Copernicus Astronomical Center, Bartycka 18, 00-716 Warsaw, Poland, \email{cheze@camk.edu.pl}
         \and 
           Department of Physics and Astronomy, University of Nigeria, Nsukka, Nigeria 
         \and 
           \c{C}anakkale Onsekiz Mart University, Faculty of Sciences, Physics Department, 17100, \c{C}anakkale, T\"{u}rkiye
         \and
           Nicolaus Copernicus Astronomical Center, Polish Academy of Sciences, ul. Rabia\'{n}ska 8, 87-100 Toru\'{n}, Poland
          }

\date{March 8, 2003}

\maketitle


\begin{abstract}
Massive stars exhibit a perplexing mismatch between their inferred masses from different observational techniques, posing a significant challenge to our understanding of stellar evolution and structure. This discrepancy is believed to be caused by the underestimation of the convective core masses. The efficiency of such measurement is usually impaired by a lot of processes at work in the interior of the stars such as convective core overshooting and interior rotation. By integrating the precision of asteroseismology which provides insights into the internal structure and dynamics of stars, with the detailed observational constraints offered by eclipsing binary systems, this study aims to precisely characterize a sample of massive stars in eclipsing binaries to infer their properties and evolutionary state. In this paper, the sample, observed photometrically with TESS and spectroscopically with SALT HRS, CHIRON, HERMES and a spectrograph at Skalnate Pleso Observatory between 2021 and 2024, are analyzed. The orbital elements as well as the basic stellar parameters of the targets in the sample are fitted to derive the geometry of their orbits as well as their absolute parameters. The asteroseismic properties of the targets are also obtained, which unravel their core dynamics and profiles.  This is a precursor work that provides detailed characterization of the targets in the sample for future theoretical modeling.

\end{abstract}

\section{Introduction}\label{intr}

There has been a perennial problem in the estimation of the masses of massive stars. The problem manifests in the mismatch between the masses inferred from model-independent observational techniques and those obtained from stellar evolution models. This was first observed by \citet{Herreroetal1992}, who pointed out that masses they obtained from spectroscopy were discrepant with masses obtained from evolutionary models. This has since been dubbed the mass discrepancy problem in massive stars and adjudged to be caused by the underestimation of the convective core mass.

Massive stars have convective cores, a radiative envelope, and a convective boundary layer between the core and the envelope. The mixing at the boundary layer transports chemical elements and, in some cases, heat to the core, thereby increasing the thermonuclear fuel in the core of the star. This, in turn, leads to an increase in the temperature and mass of the stellar core. This convective boundary mixing impairs the estimation of the convective core mass leading to the observed mass discrepancy. The boundary layer and the mixing have unfortunately proven quite difficult to calibrate. Several authors have investigated different mixing prescriptions \citep[e.g][]{Pedersenetal2021, AndersandPedersen2023} 

Several attempts have been made by several researchers to calibrate the mixing at the boundary layer and solve the  mass discrepancy problem. 
Whereas some researchers  \citep[e.g.][]{Tkachenkoetal2020} applied eclipsing binary modeling, several other researchers \citep[e.g.][]{Burssensetal2023} applied asteroseismic modeling on apparently single stars. The effects of different convective boundary mixing (CBM) mechanisms on various stellar parameters such as luminosity, radius, helium core mass e.t.c. have also been investigated \citet[e.g.][]{Johnstonetal2024}. While these efforts have improved the results obtained and our knowledge of what is happening at the boundary layer, it appears that the problem still persists. To solve this problem by leveraging on the combined strengths of eclipsing binary modeling and asteroseismology, \citet{EzeandHandler2024b, EzeandHandler2024a} searched for and discovered dozens of new $\beta$ Cephei pulsators in eclipsing binaries. To be able to use the stars in the catalog of these targets for the said purpose, their stellar parameters and properties are expected to be first determined. The paper seeks to characterize a few stars from the catalogs. It is a prerequisite work, which aims to determine the properties of the stars. These estimated parameters would serve as the input priors for the detailed modeling of the systems aimed at resolving the mass discrepancy problem.

\section{Spectroscopic Analysis}\label{sec2}

\begin{figure}[!ht]
\centering
\includegraphics[width=0.55\linewidth]{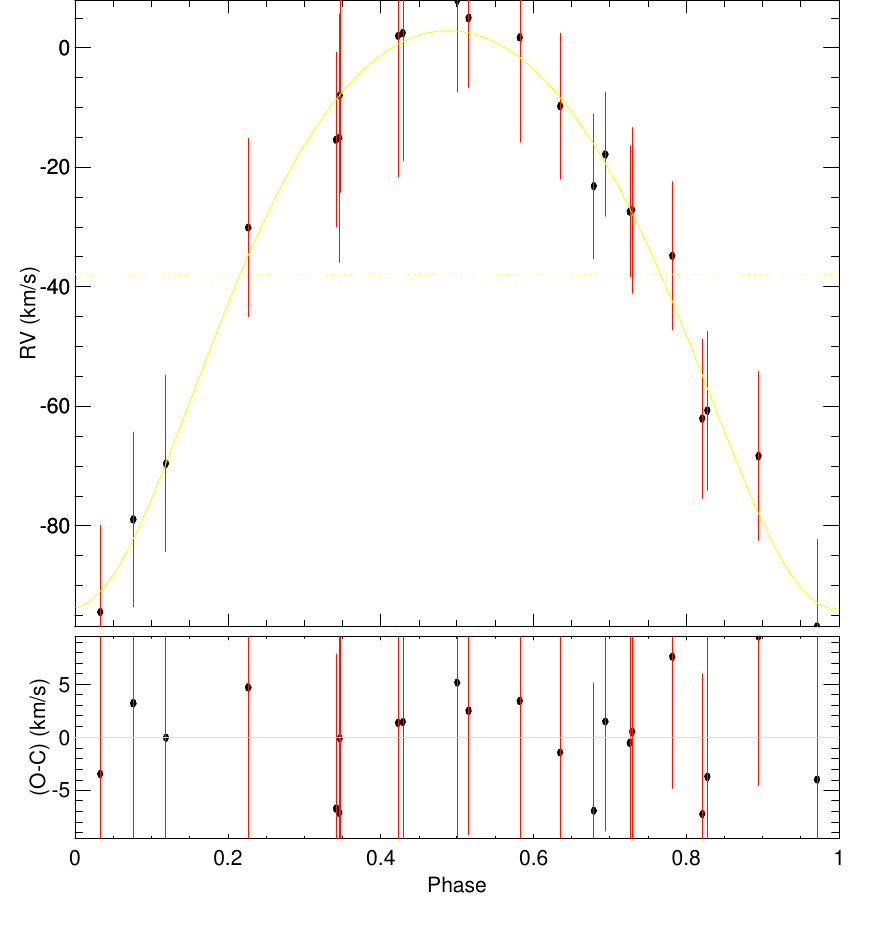}
\caption{\label{fig:Fig1}The fit to the RV curve of V1166 Cen. }
\end{figure}

Targets in \citet{EzeandHandler2024b}, which are relatively bright (brighter than V=11) with both primary and secondary eclipses visible in their photometric light curves and their pulsation spectra having several significant pulsations that appear promising for asteroseismic analysis are selected for immediate spectroscopic follow-up. This amounts to 25 targets in the sample in this work (The details are contained in our upcoming paper (Eze et al., in prep)).
Their spectra  were obtained using SALT HRS \citep{Buckleyetal2006,Bramalletal2010,Bramalletal2012,Crauseetal2014,Crawfordetal2010}, CHIRON spectrograph in SMARTS 1.5-m  telescope operated by the SMARTS Consortium \citep{Tokovininetal2013, Paredesetal2021}, HERMES spectrograph \citep{Raskinetal2011}  and the MUSICOS-type spectrograph at 1.3-m telescope of Skalnate Pleso observatory (SPO) \citep{BaudrandandBohm1992, Pribullaetal2024} using appropriate exposure time, resolution, signal-to-noise ratio and under suitable sky conditions. The observed spectra were normalized and the radial velocities (RV) as well as the atmospheric solutions of the targets were determined. The RVs were extracted using the IRAF/PYRAF FXCOR task \citep{Tody1986,2012ascl.soft07011S} in cross-correlation with synthetic spectra obtained with the Spectrum code \citep{Gray1999} using ATLAS model atmospheres \citep{CastelliandKurucz2003}. Appropriate values of $T_{eff}$,  logg and vsini used for the synthetic spectra of each target were inferred from preliminary fits to the spectra using HANDY\footnote{https://github.com/RozanskiT/HANDY)}. The RV curves were fitted using rvfit \citep{Iglesiasetal2015} to obtain the orbital elements/parameters of the systems. Figure \ref{fig:Fig1} shows the fitted RV curve and Table \ref{table:Tab1} gives the fitted parameters of one of the targets (V1166 Cen) in the sample. V1166 Cen is used as an example target in this paper to demonstrate the analysis where applicable.

In the case of V1166 Cen, all the orbital parameters are fitted as free parameters. It has an eccentric orbit with eccentricity of $0.2(1)$, and a radial velocity semi-amplitude ($K$) of $48(5)\,km\,s^{-1}$. The mass function of the system  is $0.15(8)$\,$M_\odot$. The period agrees with the photometric period reported in \citet{EzeandHandler2024b} up to 1 decimal places.

\begin{center}
{\scriptsize
\begin{table}[ht]
\centering
\caption{The fitted parameters of V1166 Cen.} \label{table:Tab1}
\begin{tabular}{lr}
\multicolumn{2}{c}{}\\
\hline
\hline
Parameter			&Value\\
\hline
\multicolumn{2}{c}{Adjusted Quantities}\\
\hline
$P$ (d)		& 13.6(2)\\
$T_p$ (HJD)		& 2459956(1)\\
$\textbf{e}$			& 0.2(1)\\
$\omega$ (deg)		& 183(34)\\
$\gamma$ (km/s)	& -37(3)\\
$K_1$ (km/s)		& 48(5)\\
\hline
\multicolumn{2}{c}{Derived Quantities}\\
\hline
$a_1\sin i$ ($10^6$ km)	& 9(1)\\
$f(m_1,m_2)$ ($M_\odot$)	& 0.15(8)\\
\hline
\multicolumn{2}{c}{Other Quantities}\\
\hline
$\chi^2$		& 2.39\\
$N_{obs}$ (primary)	& 22\\
Time span (days)	& 73.06\\
$rms_1$ (km/s)	& 4.64\\
\hline
\end{tabular}
\end{table}
}
\end{center}

The atmospheric parameters ($T_{eff}$, logg and vsini) were also obtained by cross-matching the normalized observed spectrum with a synthetic model atmosphere template obtained using the Non-Local Thermodynamic Equilibrium (NLTE) Tlusty BSTAR2006 grids \citep{HubenyandLanz2017}. For the $T_{eff}$ and $\log g$, $H_{\beta}$ or $H_{\gamma}$ was used whereas HeI 4387 and/or HeI 5878 lines were used for $v \sin i$. The errors obtained are at 1-sigma ($1{\sigma}$) tolerance in the goodness of fit parameter. For example, the $T_{eff}$, $\log g$ and $v \sin i$ of V1166 Cen are 27000(1000)\,K, 3.8(1)\,dex and 252(10)\,km\,s$^{-1}$  respectively. The details of the analysis are contained in our upcoming paper (Eze et al., in prep), which discusses the spectroscopic follow-up on a sample of $\beta$ Cephei pulsators in eclipsing binaries. 

For double-lined spectroscopic binaries in our sample, the spectra are first disentangled using the FD3 program (\citep{Ilijicetal2004} and the atmospheric parameters of each of the individual components estimated from the disentangled spectra. FD3 is a C-based program, which performs the separation of the spectra of double-lined spectroscopic binaries in the Fourier space. An example of the double-lined spectroscopic binaries in our sample is V1216 Sco. The orbital and atmospheric solutions of this system, for instance, are used as priors  for the detailed binary modeling (Miszuda et al., submitted to $A\&A$).

\section{Pulsation Analysis}\label{sec3}
\begin{figure}[!ht]
\centering
\includegraphics[width=0.8\linewidth]{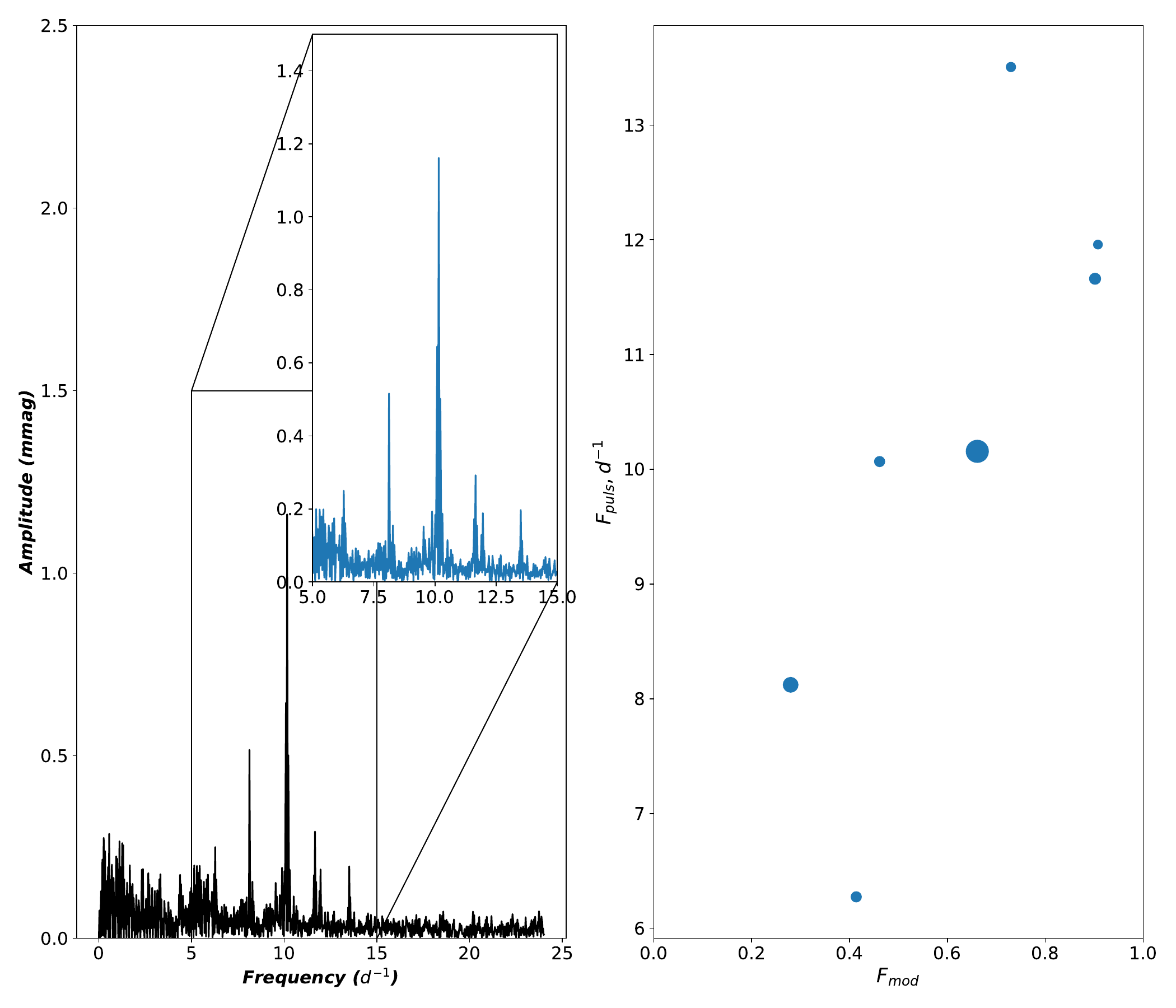}
\caption{Fourier spectra of V1166 Cen with the neighborhood of the dominant pulsation zoomed into (left panel) and \'{e}chelle diagram of V1166 Cen (right panel). \label{fig:Fig2} }
\end{figure}

The preliminary pulsation analyses were done as described in detail, in \citet{EzeandHandler2024b}. Additionally, at this time, all significant $\beta$ Cephei pulsations in the systems are extracted instead of a few dominant pulsations. Figure \ref{fig:Fig2} (left panel) and Table \ref{table:Tab2} show the Fourier spectrum and pulsation frequencies respectively of V1166 Cen. Figure \ref{fig:Fig2} (left panel) shows a possible rotationally split quadrupole mode of l=2. There is an observed asymmetry in the split mode. Asymmetries in rotationally split mode multiplets were recently comprehensively discussed by \citet{Guoetal2024} who examined the effects of perturbations from rotational non-spherical distortions, intrinsic magnetic fields, resonant mode coupling, and binarity on the observed frequency splittings. V1166 Cen is a fast rotator as suggested by its $v \sin i = 252(10)$\,km\,s$^{-1}$ and the asymmetry is most likely to originate from the rotationally induced non-spherical distortions.     

Further pulsation analysis was also conducted by plotting the \'{e}chelle diagrams of the targets using all significant frequencies found in the data. Figure \ref{fig:Fig2} (right panel) shows the \'{e}chelle diagram of V1166 Cen. There are observed curved patterns in the \'{e}chelle diagram, which likely arise from the effects of rotation.  
The exact mode characteristics ($\ell$, $m$, $n$) of the systems in our  sample are yet to be confirmed. The modes are inferred via rotational splitting or the use of ground-based multicolor photometric data. For systems where the ground-based multicolor photometric follow-up are feasible, the multicolor photometric observations have been conducted with SHOC in the 1.0 m SAAO telescope \citep{Coppejansetal2013} and are currently being reduced. For systems like V1166 Cen where the amplitude of even its strongest mode is too small to reliably conduct actual mode identification with ground-based multicolor photomeric data without excessive usage of telescope time, only the rotational splitting method has been used and no ground-based multicolor photometric follow up is planned. 

\begin{table}[]
    \centering
    \caption{The Pulsation Frequencies of V1166 Cen. 'F' is the frequency and 'A' is the amplitude of the frequency. 
    \label{table:Tab2}}
    \begin{tabular}[ht!]{lll}
    \hline\hline
&F\,($d^{-1}$)&A\,(mmag) \\ 
\hline
F1&	10.157(1)&	 1.17(6)  \\
F2&	8.122(3)&	 0.52(6)  \\
F3&	11.661(5)&	 0.29(6) \\
F4&	6.274(6)&	 0.24(6)  \\
F5&	10.068(6)&	 0.24(6) \\
F6&	11.959(8)&	 0.18(6)  \\
F7&	13.506(7)&	 0.20(6) \\
\hline
    \end{tabular}
\end{table}

\section{Ensemble Pulsational Characteristics}\label{sec4}
\begin{figure}[!ht]
\centering
\includegraphics[width=0.8\linewidth]{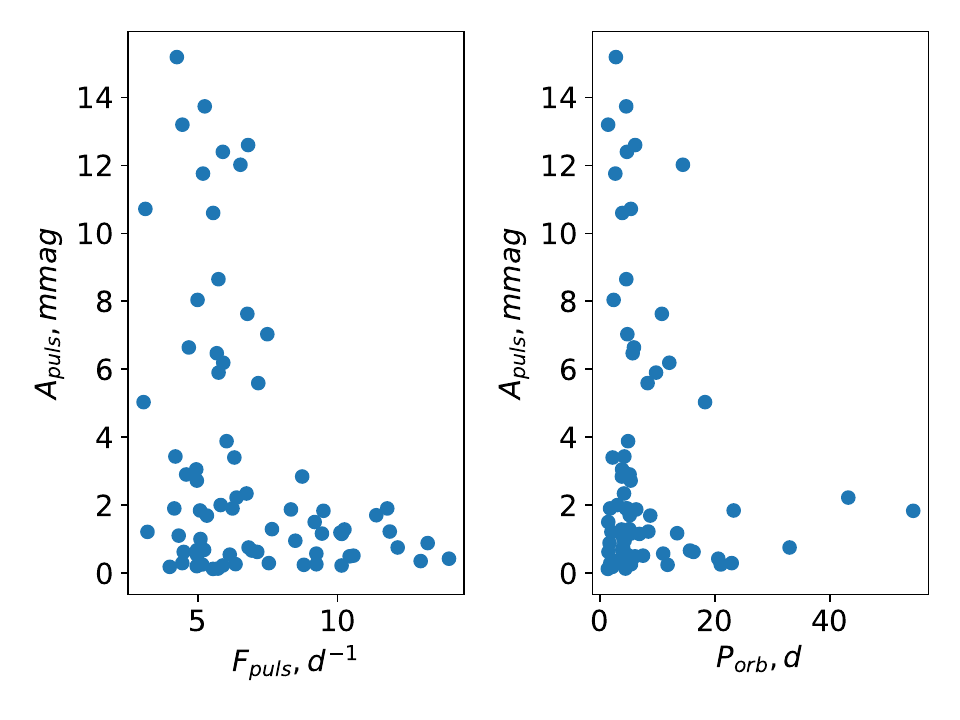}
\caption{\label{fig:Fig3} Frequency-amplitude relation for $\beta$ Cep p modes (left panel) and Period-pulsation amplitude relation for our sample of $\beta$ Cep pulsators in eclipsing binaries (right panel).}
\end{figure}

With the now considerably larger sample of $\beta$ Cephei stars in eclipsing binaries \citep[see][]{EzeandHandler2024b}, the ensemble pulsational characteristics of $\beta$ Cephei pulsators in eclipsing binaries are investigated. We plotted different pulsational and orbital parameters obtained in \citet{EzeandHandler2024b} for the sample reported in \citet{EzeandHandler2024b}. The plot of amplitude against the dominant pulsation frequency in Figure \ref{fig:Fig3} (left panel) shows that the amplitude of p-mode pulsations decays exponentially with an increase in pulsation frequency. One outlier with  pulsation amplitude of $116.99(9)\,mmag$ was removed from the plots to avoid its undue driving force to the trends.  This appears to be in line with \citet{Fritzewskietal2024}, who observed a similar decay in the amplitude of $\beta$ Cephei pulsations.

To check the impact of the orbit or possible tidal interactions of the systems on the amplitude of pulsations in eclipsing binaries, the amplitude of the pulsation is plotted against the orbital period as shown in Figure \ref{fig:Fig3} (right panel). The preliminary result shows an exponential decay. It is not yet clear whether the closeness or possible tidal interactions as a result, enhance or dampen the amplitude of the pulsations. This is owing to the fact that the sample was treated, at this point, as a perfectly homogeneous sample without distinction between systems where tidal interactions are possible and where they are not. A possible selection effect is not ruled out as there are much less systems with long orbital periods. A detailed ensemble analysis is underway to explore the possible orbital period-pulsation frequency and orbital period-pulsation amplitude relations, among others,  in $\beta$ Cephei stars in eclipsing binary systems. Such a detailed study has been conducted for $\delta$ Scuti stars in eclipsing binaries by \citet{LiakosandNiarchos2017}, who observed a linear $\log P_{pul} - \log P_{orb}$ relation for stars with orbital period $< 13$\,d. However, preliminary results do not show such a relation in $\beta$ Cephei stars in eclipsing binaries.

\section{Conclusion}\label{sec5}
A sample of $\beta$ Cephei stars in eclipsing binaries is characterized to obtain their stellar properties. Their orbital as well as their atmospheric solutions are determined. The $T_{eff}$, $\log g$ and $v \sin i$ of V1166 Cen, for instance, are 27000(1000)\,K, 3.8(1)\,dex and 252(10)\,km\,s$^{-1}$  respectively. The pulsational mode characteristics that are visible in the Fourier spectra and \'{e}chelle diagrams of the stars in our sample are investigated. The ensemble pulsational characteristics are also reported. Multicolor photometry for  direct mode identification is currently being reduced.


\acknowledgements
The work is supported by the Polish National Science Foundation (NCN) under grant nr 2021/43/B/ST9/02972.
C.I. Eze is grateful for the support received from the staff of the Stellar Department, Astronomical Institute of the Slovak Academy of Sciences during his observation runs at the Skalnate Pleso Observatory.

\bibliographystyle{caosp309}
\bibliography{caosp309}
\end{document}